\documentstyle[sprocl]{article}

\input{psfig}

\def\be{\begin{equation}}
\def\ee{\end{equation}}
\def\bea{\begin{eqnarray}}
\def\eea{\end{eqnarray}}
\def\Ref#1{(\ref{#1})}
\def\cropen#1{\crcr\noalign{\vskip #1}}
\def\crr{\cropen{1\jot }}
\def\Red{}
\def\Black{}

\def\cA{{\cal A}}

\def\plushc{\ +\ h.c.}

\newcommand{\I}{{\rm i}}

\newcommand{\cL}{{\cal L}}
\newcommand{\cD}{{\cal D}}
\newcommand{\cF}{{\cal F}}

\newcommand{\nn}{\nonumber}

\catcode`@=11
\def\citenum#1{\csname b@#1\endcsname}
\catcode`@=12

\begin{document}

\title{NEW SUPERSYMMETRY ALGEBRAS FROM\\ PARTIAL SUPERSYMMETRY BREAKING}

\author{JONATHAN A.~BAGGER \\ RICHARD ALTENDORFER}

\address{Department of Physics and Astronomy\\
Johns Hopkins University\\
Baltimore, MD 21218, USA} 


\maketitle\abstracts{In this talk we will study the partial breaking
of supersymmetry in flat and anti de Sitter space.  We will see that
partial breaking in flat space can be accomplished using either of
two representations for the massive $N=1$ spin-3/2 multiplet.  We
will ``unHiggs" each representation and find a new $N=2$ supergravity
and a new $N=2$ supersymmetry algebra.  We will also see that partial
supersymmetry breaking in AdS space can give rise to a new $N=2$
supersymmetry algebra, one that is necessarily nonlinearly
realized.}

\section{Introduction}

It is a sad fact of life that modern particle physicists can be
classified by an integer, $N$, which counts the number of
supersymmetries they assume to be active in the physical world.
String theorists, for example, live and work in a rarefied region 
where $N=8$ supersymmetry appears to hold sway.
Most experimentalists, on the other hand, toil in great laboratories
where $N=0$ supersymmetry rules the day.  In between are the
phenomenologists, who are busy preparing for the time when
experimentalists will study the first supersymmetry, that of $N=1$.

This hierarchy of supersymmetry is not only a sociological fact but
a physical necessity as well.  From string theory we know that the
real world has $N=8$ supersymmetry.  These supersymmetries must be
spontaneously broken, either all at once, to $N=0$, or partially,
first to $N=1$ (or higher), and then to $N=0$.  The spontaneous breaking
of extended supersymmetry, from $N=8$ to $N=1$ to $N=0$, is what ties
together the different regions of the physical world.

The hierarchy of supersymmetries can be described phenomenologically
using the language of effective field theory.  In particle physics,
this language was first developed during the 1960's to describe the
chiral symmetry breaking associated with pions, protons, and neutrons.
By now the formalism has been sufficiently well developed that it can
be used as a framework for understanding all of field theory,
especially the physics associated with spontaneous symmetry breaking.

From this point of view, there is an ultimate theory, perhaps M
theory, that exists at high energies.  At each lower energy, one
integrates out the degrees of freedom associated with the high
energies, and constructs a nonrenormalizable, effective field theory.
The effective field theory contains only those degrees of freedom
that are relevant for physics at the scale under study.  In the
effective field theory, unbroken symmetries are realized linearly,
while spontaneously broken symmetries are realized nonlinearly.
The nonlinear symmetries are the remnants of symmetries spontaneously
broken at higher scales.

In the context of supersymmetry, we are interested in the effective
field theory that comes from breaking $N=8$ down to $N=1$.  In this
talk, for simplicity, we will focus our attention on the easiest
case, of $N=2$ broken to $N=1$.  We will construct effective
theories that contain
an unbroken, linearly realized $N=1$ supersymmetry, together with a
spontaneously broken, nonlinearly realized, $N=2$.

At first glance, it might seem impossible to partially break $N=2$
to $N=1$.  The argument runs as follows.  Start with the $N=2$
supersymmetry algebra,
\bea
\{ Q_\alpha,\,\bar
Q_{\dot\alpha} \} &\ =\ & 2\, \sigma^m_{
\alpha\dot\alpha}\,P_m \nonumber \\
\{ S_\alpha,\,\bar
S_{\dot\alpha} \} &=& 2\, \sigma^m_{
\alpha\dot\alpha}\,P_m\ ,
\eea
where $Q_\alpha$ and its conjugate $\bar Q_{\dot\alpha}$ denote the
first, unbroken supersymmetry, and $S_\alpha$, $\bar S_{\dot\alpha}$
the second.  Suppose that one supersymmetry is not broken, so
\be
Q\, |0\rangle \ =\ \bar Q\, |0\rangle\ =\ 0\ .
\ee
Because of the supersymmetry algebra, this implies that the Hamiltonian
also annihilates the vacuum,
\be
H\, |0\rangle \ =\ 0\ .
\ee
Then, according to the supersymmetry algebra,
\be
(\bar S S + S \bar S)\, |0\rangle\ =\ 0\ .
\label{SSbar}
\ee
The final step is to peel apart this relation and conclude that
\be
S\, |0\rangle\ =\ \bar S\, |0\rangle\ =\ 0\ .
\ee
From this line of reasoning, one might think that partial breaking
is impossible.

Fortunately, this argument has two significant loopholes.  The first is that,
technically-speaking, spontaneously-broken charges do not exist.  Indeed,
in a spontaneously broken theory, one only has the right to consider
the algebra of the {\it currents.}  For the case at hand, the current
algebra can be modified as follows,
\bea
\{ \bar Q_{\dot\alpha},\
J^1_{\alpha m}
\} &\ =\ & 2 \,\sigma^n_{\alpha\dot\alpha}\,
T_{mn}\nonumber \\
\{ \bar S_{\dot\alpha},\ J^2_{\alpha m}
\} &=& 2\, \sigma^n_{\alpha\dot\alpha}\,
(v^4 \eta_{mn}+ T_{mn})\ ,
\eea
where the $J^i_{\alpha m}$ ($i = 1,2$) are the supercurrents and $T_{mn}$
is the stress-energy tensor.  Note that Lorentz invariance does not
force the right-hand sides of the commutators to be the same.  If there
were no first supersymmetry, the $v^4$ term in the second commutator
could be absorbed in $T_{mn}$; it would play the role of the vacuum
energy.  However, the first supersymmetry can be said to {\it define}
the stress-energy tensor, in which case there is an extra term in the
second commutator.  This discrepancy
prevents the current algebra from being integrated into a charge algebra,
so the no-go theorem is avoided.

The second loophole involves the last step of the theorem.  Even if the
supercharges were to exist, it is only possible to extract (5) from (4)
if the Hilbert space is positive definite.  In covariantly-quantized
supergravity theories, this is not the case: the gravitino $\psi_{m\alpha}$
is a gauge field with negative-norm components.

There are, by now, many examples of partial supersymmetry breaking
which exploit the first loophole.  The first was
given by Hughes, Liu and Polchinski,\cite{hlp} who showed that supersymmetry
is partially broken on the world volume of an $N=1$ supersymmetric 3-brane
traveling in six-dimensional superspace.  Later, Bagger and
Galperin\cite{2bagger,BGT}
used the techniques of Coleman, Wess, Zumino,\cite{cwz} and 
Volkov\cite{volkov} to
construct an effective field theory of partial supersymmetry breaking,
with the broken supersymmetry realized nonlinearly.  They found that the
Goldstone fermion could belong to an $N=1$ chiral {\it or} an $N=1$
vector multiplet.  At about the same time, Antoniadis, Partouche and
Taylor discovered another realization in which the Goldstone fermion
is contained in an $N=2$ vector multiplet.\cite{apt}

These results leave open many important questions.  First and foremost,
one would like to know how partial breaking works in the presence of
gravity.  Gravity couples to the true stress-energy tensor, so it
distinguishes between the right-hand sides of the commutators (6).  
Some early work on this question was done by Cecotti,
Girardello and Porrati,\cite{cgp} and by Zinov'ev.\cite{zin}  (A geometrical
interpretation was given in Ref.~9.)  These groups considered
nonminimal cases and found that their gravitational couplings utilize
the second loophole.  One would like to reconcile these results with
those above.  For reasons that will soon become clear, one would also
like to know how partial breaking works in the presence of a nonvanishing
cosmological constant.

In this talk we will address these questions using nonlinear realizations.
The theory of nonlinear realizations provides a minimal,
model-independent approach to the questions associated with partial
supersymmetry breaking.  We will focus on two of the multiplets
of Bagger and Galperin,\cite{2bagger} and couple each of them to supergravity, to
lowest nontrivial order.

During the course of our work, we will find that partial breaking in
flat space motivates an alternative representation for the $N=1$
massive spin-3/2 multiplet.  When coupled to gravity, this representation
gives rise to a new $N=2$ supergravity with a different $N=2$ supersymmetry
algebra.  We shall also see that partial breaking in anti de Sitter
space can give rise to a new $N=2$ supersymmetry algebra.

In each case, our technique will be as follows:  We will start by
constructing the Lagrangian and supersymmetry transformations for the
massive $N=1$ spin-3/2 multiplet.  We shall then ``unHiggs" the
representation by adding appropriate Goldstone fields and coupling
gravity.  We will see that the basic technique works in
flat and AdS space.

\section{The Massive $N=1$ Spin-3/2 Multiplet in Flat Space}

The starting point for our investigation is the massive $N=1$
spin-3/2 multiplet.  This multiplet contains six bosonic and
six fermionic degrees of freedom, arranged in states of the
following spins,
\be
\pmatrix{
{3\over2} \crr
1\ \ \ 1 \crr
{1\over2}}\ .
\ee
The traditional representation of this multiplet contains
the following fields: \cite{fvn}  one spin-3/2 fermion, one spin-1/2 fermion,
and two spin-one vectors, each of mass $m$.  The alternative
representation has the same fermions, but just one vector plus
one antisymmetric tensor.\cite{ogsok}  As we shall see, each representation
has a role to play in the theory of partial supersymmetry breaking.

The traditional representation is described by the following
Lagrangian,\cite{fvn}
\bea
\cL &\ = \ & \epsilon^{m n \rho \sigma} \overline \psi_{m}
  \overline \sigma_n \partial_\rho \psi_\sigma 
 - \I \overline \zeta \overline \sigma^m \partial_m \zeta
 - {1 \over 4} \cA_{m n} \bar\cA^{m n} \nonumber \\
& &-\ {1\over 2}m^2\, \cA_m \bar\cA^m  
\ +\ {1\over 2}m\,\zeta\zeta  \ +\ {1\over 2}m\,\bar\zeta\bar\zeta \nonumber \\[1mm]
& & -\ m\,\psi_m \sigma^{m n} \psi_n -\ m\,\bar\psi_m \bar\sigma^{m n} \bar\psi_n\ .
\eea
Here $\psi_m$ is a spin-3/2 Rarita-Schwinger field, $\zeta$ a spin-1/2
fermion, and $\cA_m = A_m + \I B_m$ a complex spin-one vector.  This Lagrangian
is invariant under the following $N=1$ supersymmetry transformations,
\bea
\delta_\eta \cA_m &\ =\ & 2\psi_m\eta - \I{2\over\sqrt{3}}
\bar\zeta\bar\sigma_m\eta 
-{2\over \sqrt{3}m}\partial_m(\zeta\eta) \nonumber \\
\delta_\eta \zeta &=& {1\over\sqrt{3}}\bar\cA_{mn}\sigma^{mn}
\eta -\I{m\over\sqrt{3}}
\sigma^m\bar\eta \cA_m \nonumber \\
\delta_\eta \psi_m &=& {1\over 3m}\partial_m(\bar\cA_{rs}
\sigma^{rs} \eta + 2\I m 
\sigma^n\bar\eta \cA_n)  
- {\I\over 2}(H_{+mn}\sigma^n + {1\over 3}H_{-mn}\sigma^n)
\bar\eta \nonumber
\\ & & -\ {2\over 3}m({\sigma_m}^n \bar\cA_n \eta + \bar
\cA_m\eta)\ ,
\eea
where $H_{\pm mn}=\cA_{mn}\pm {\I\over 2}\epsilon_{mnrs}\cA^{rs}$.

The alternative representation has the following Lagrangian,
\bea
\cL &\ =\ & \epsilon^{pqrs} \bar \psi_{p}
  \bar \sigma_q \partial_r \psi_s
 - \I \bar \zeta \bar \sigma^m \partial_m \zeta
 - {1 \over 4} A_{mn} A^{mn} 
 + {1\over 2}v^m v_m \nonumber \\
& & -\ {1\over 2}m^2 A_m A^m  - {1\over4}m^2
B_{mn}B^{mn} \ +\ {1\over 2}m\,\zeta\zeta  \ +\ {1\over
2}m\,\bar\zeta\bar\zeta\nonumber\\[1mm]
& & -\ m\,\psi_m \sigma^{mn} \psi_n \ -\ m\,\bar\psi_m \bar\sigma^{mn}
\bar\psi_n\ ,
\eea
where $A_{mn}$ is the field strength associated with the real vector
field $A_m$, and $v_m = {1\over 2}\epsilon_{mnrs} \partial^n B^{rs}$ is
the field strength for the antisymmetric tensor
$B_{mn}$.  This Lagrangian is invariant under the following
$N=1$ supersymmetry transformations,
\bea
\delta_\eta A_m &\ =\ & (\psi_m\eta + \bar\psi_m\bar\eta) + 
{\I\over\sqrt{3}}
(\bar\eta\bar\sigma_m\zeta - \bar\zeta\bar\sigma_m\eta)
-{1\over \sqrt{3}m}\partial_m(\zeta\eta +  \bar\zeta\bar\eta) 
\nonumber \\
\delta_\eta B_{mn} &=& {2\over\sqrt{3}}\left( \eta\sigma_{mn}\zeta 
+ {\I\over 2m}\partial_{[m}\bar\zeta\bar\sigma_{n]}\eta \right) 
\ +\ \I\eta\sigma_{[ m}\bar\psi_{n ]} + {1\over
m} \eta\psi_{mn}
\plushc \nonumber\\
\delta_\eta \zeta &=& {1\over\sqrt{3}} A_{mn}\sigma^{mn}\eta - 
{\I m\over\sqrt{3}}
 \sigma^m\bar\eta A_m - {1\over\sqrt{3}}m\sigma_{mn}\eta B^{mn} - 
{1\over\sqrt{3}} v_m\sigma^m\bar\eta \nonumber \\
\delta_\eta \psi_m &=& {1\over 3m}\partial_m \left( A_{rs}
\sigma^{rs}\eta + 2\I m
\sigma^n\bar\eta A_n  \right) 
 -{\I\over 2} (H^A_{+mn}\sigma^n + {1\over 3}H^A_{-mn}\sigma^n)
\bar\eta \nonumber \\
& & -\ {2\over 3}m ({\sigma_m}^n A_n \eta +
A_m\eta)\ +\ {1\over 3m}\partial_m \left( 2
v_n\sigma^n\bar\eta  - m 
 \sigma^{rs}\eta B_{rs} \right) \nonumber\\
&&    -\ {2\I\over 3} (v_m + \sigma_{mn}v^n)\eta 
 - {\I m\over 3} (B_{mn}\sigma^n\bar\eta + \I
\epsilon_{mnrs}B^{n r}\sigma^s\bar\eta) \ ,
\eea
where the square brackets denote antisymmetrization, without a
factor of 1/2.

These Lagrangians describe the free dynamics of massive spin-3/2
and 1/2 fermions, together with their supersymmetric partners,
massive spin-one vector and tensor fields.  They can be thought
of as ``unitary gauge" representations of theories with additional
symmetries:  a second supersymmetry for the massive spin-3/2 fermion,
and additional gauge symmetries associated with the massive gauge
fields.

To study partial breaking, we need to ``unHiggs" these Lagrangians
by including appropriate gauge and Goldstone fields.  In each case
we will need to add a Goldstone fermion and gauge the full $N=2$
supersymmetry.  The supersymmetric partners of the fermion will
turn out to be the Goldstone bosons that restore
the gauge symmetries associated with the massive bosonic fields.
In this way we will
construct two theories with $N=2$ supersymmetry nonlinearly realized,
and $N=1$ represented linearly on the fields. The resulting effective
field theories describe the physics of partial supersymmetry breaking,
well below the scale where the second supersymmetry is broken.

The trick to this construction is to add the right fields.  Because
$N=1$ supersymmetry
is not broken, the Goldstone fermion must belong to an $N=1$
supersymmetry multiplet.  For the two cases of interest, we shall
see that the Goldstone
fermion must be in a chiral {\it or} a vector multiplet.

Let us first consider the chiral case.  Under the first supersymmetry,
a complex boson $\phi$ transforms into a Weyl fermion $\chi$,
\be
\delta_{\eta^1}\phi \ =\ \sqrt 2\,\eta^1\chi\ .
\ee
If $\chi$ is the Goldstone fermion, it shifts under the second
supersymmetry,
\be
\delta_{\eta^2}\chi\ =\ \sqrt 2\,v^2\,\eta^2\ +\ \ldots\ ,
\ee
where $v$ is the scale of the second supersymmetry breaking.  Therefore
the closure of the two supersymmetries on $\phi$ gives
\be
[\,\delta_{\eta^2},\,\delta_{\eta^1}\,]\,\phi\ =\ 2\,v^2\,\eta^1\eta^2\ +
\ \ldots
\ee
We see that the complex scalar $\phi$ undergoes a constant shift.  This implies
that $\phi$ itself is a Goldstone boson.  It expects to be eaten by a
complex vector field, which suggests that the chiral Goldstone multiplet
should be associated with the traditional representation for the massive
spin-3/2 multiplet.

As shown in Figure 1(a), the degree of freedom counting works out just right.
We start with the $N=1$ chiral Goldstone multiplet and add an $N=1$
vector multiplet.  These fields may be thought of as $N=1$ matter.  We then add
the gauge fields of $N=2$ supergravity.  As we will see, the
full set of fields can
be used to construct a Lagrangian which is invariant under $N=2$ supersymmetry.
The results look complicated, but they are actually very simple:  In unitary
gauge, the two vectors eat the two scalars, while the Rarita-Schwinger field
eats one linear combination of the spin-1/2 fermions.  This leaves the
massive $N=1$ multiplet coupled to $N=1$ supergravity.

\begin{figure}[t]
\hspace*{0.1truein}
\psfig{figure=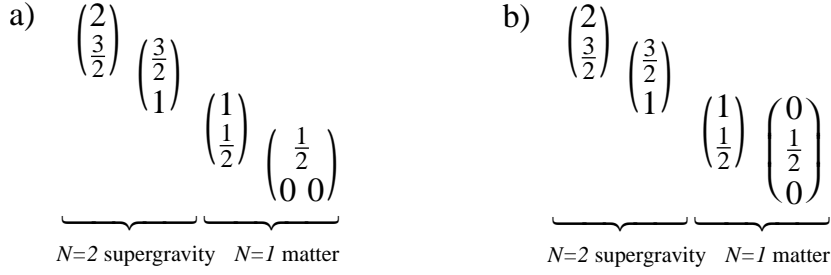,height=1.4in}
\caption{The unHiggsed versions of the (a) traditional and (b)
alternative representations of the $N=1$ massive spin-3/2 multiplet.}
\end{figure}

With that said, we now present the Lagrangian:
\bea
&& e^{-1}\cL \ =\ \nn\\
&& -\ {1 \over 2 \kappa^2} {\cal R}
 + \epsilon^{m n r s} \overline \psi_{m i}
  \overline \sigma_n D_r \psi^i_s 
 - \I \overline \chi \  \overline \sigma^m D_m \chi 
 - \I \overline \lambda \overline \sigma^m D_m \lambda
-  \cD^m \phi \overline{\cD_m \phi} \nn\\
&& -\ {1 \over 4} \cA_{m n} \overline \cA^{m n} 
 - \ \Bigl( {1 \over \sqrt{2}} m  \psi^2_m \sigma^m
   \overline \lambda 
 + \I m\psi^2_m \sigma^m
   \overline \chi 
+ \sqrt{2}  \I m \lambda \chi 
 +\ {1 \over 2} m \chi \chi \nn\\
&& +\ \Red m \Black\, \psi^2_m \sigma^{m n} \psi^2_n 
+\ {\kappa \over 4} \epsilon_{i j}
\psi^i_m \psi_{n}^{j}
    \overline H_+^{m n} 
 + {\kappa \over \sqrt{2}}  \chi \sigma^m
    \overline \sigma^n \psi^1_m \overline{\cD_n \phi} 
 \nonumber \\
& & + \  {\kappa \over 2 \sqrt{2}}  \overline
\lambda
    \overline \sigma_m \psi^1_n \overline H_-^{m n}
 +  {\kappa \over \sqrt{2}}
    \epsilon^{m n r s} 
    \overline \psi_{m 2} \overline \sigma_n \psi^1_r
    \overline{\cD_s \phi}
    \plushc \Bigr) \ ,
\eea
where $\kappa$ denotes Newton's constant, $m = \kappa v^2$, and
\bea
\cA_m&\ =\ &A_m + \I B_m \nonumber \\
\cA_{mn}&=&\partial_m \cA_m - \partial_n \cA_m \nonumber\\
H_{\pm mn}&=&\cA_{mn}\pm {\I\over
2}\epsilon_{mnrs}\cA^{rs}\ .
\eea
The supercovariant derivatives are as follows,
\bea
\hat\cD_m\phi&\ =\ &\partial_m\phi - {\kappa\over\sqrt{2}}\psi^1_m\chi - 
{1\over\sqrt{2}} \kappa v^2  \cA_m\nonumber \\
\hat\cA_{mn}&=&\cA_{mn} + \kappa\psi^2_{ [m}\psi^1_{n] } - 
{\kappa\over\sqrt{2}}\bar\lambda\bar\sigma_{ [n}\psi^1_{m ]} \ .
\eea
This Lagrangian is invariant under two independent abelian gauge symmetries,
as well as the following supersymmetry transformations,
\bea
\delta e^a_m &\ =\ &\I \kappa (\eta^i \sigma^a \overline \psi_{m i} + 
   \bar\eta_i \bar\sigma^a \psi_{m}^i) \nonumber \\
\delta \psi^i_{m} & = & {2 \over \kappa} D_m \eta^i
     \nonumber \\
& &   +\ \left( -{\I \over 2} \hat H_{+m n}
\sigma^n \overline \eta_1
 + \sqrt{2} \overline{\cD_m \phi} \eta^1 
 - \kappa\psi^1_m(\bar\chi\bar\eta_1) 
 + \I v^2 \sigma_m \overline \eta_2 \right){\delta_2}^i \nonumber \\
\delta \cA_m &=& 2 \epsilon_{i j} \psi_{m}^{i} \eta^j 
+ \sqrt{2}  \overline \lambda \overline \sigma_m \eta^1 \nonumber \\
\delta \lambda  &=&  {\I \over \sqrt{2}} \overline{\hat\cA}_{mn} 
\sigma^{mn} \eta^1
\Red - \I \sqrt{2} v^2 \eta^2 \Black \nonumber \\
\delta \chi  &=&  \I \sqrt{2} \sigma^m {\hat\cD_m \phi}\overline \eta_1
\Red + 2 v^2 \eta^2 \Black \nonumber \\
\delta \phi &=& \sqrt{2} \chi \eta^1 \ ,
\eea
for $i=1,2$.  This result holds to leading order, that is, up
to and including terms in the transformations that are linear in
the fields.  Note that this representation is irreducible in the sense that
there are no subsets of fields that transform only into themselves
under the supersymmetry transformations.

Let us now consider the vector case.  Under the first supersymmetry,
the real vector $B_m$ of a vector multiplet 
transforms into a Weyl fermion $\lambda$,
\be
\delta_{\eta^1} B_m\ =\ \sqrt 2\I\,(\lambda\sigma_m\bar\eta^1 -
\eta^1\sigma_m\bar\lambda)\ .
\ee
If $\lambda$ is the Goldstone fermion, it shifts under the second
supersymmetry.  Therefore the closure of the two supersymmetries
on $B_m$ gives
\be
[\,\delta_{\eta^2},\,\delta_{\eta^1}\,]\,B_m\ =\ 2\I v^2\,(\eta^2\sigma_m\bar\eta^1 -
\eta^1\sigma_m\bar\eta^2)\ +
\ \ldots
\ee
From this we see that the real vector $B_m$ is a Goldstone
boson.  It expects to be eaten by an antisymmetric tensor field.
This suggests that the vector Goldstone multiplet should
be associated with the alternative representation for the
massive spin-3/2 multiplet.

The degree of freedom counting is shown in Figure 1(b).  As before,
we include the $N=2$ supergravity multiplet.  This time, however,
the matter fields include the $N=1$ vector Goldstone multiplet,
together with one $N=1$ tensor multiplet.  In unitary gauge, one
vector eats one scalar, while the antisymmetric tensor eats the
other vector.  These are the minimal set of fields that arise
when coupling the alternative spin-3/2 multiplet to $N=2$
supergravity.

The Lagrangian for this system can be worked out following the
same procedure described above.  We find
\bea
& &e^{-1}\cL\ =\ \nonumber \\
&  &  -\ {1 \over 2 \kappa^2} {\cal R}
 + \epsilon^{pqrs} \bar \psi_{p i}
  \bar \sigma_q D_r \psi^i_s 
 - \I \bar \chi \bar \sigma^m D_m \chi 
 - \I \bar \lambda \bar \sigma^m D_m \lambda 
-  {1\over 2}\cD^m \phi \cD_m \phi \nn\\
&& -\ {1 \over 4} \cF^A_{mn} \cF^{Amn} - {1\over4}\cF^B_{mn}\cF^{Bmn}
+ {1\over 2}v^m v_m
  - \Bigl( {1 \over \sqrt{2}}  m \,  \psi^2_m \sigma^m
   \bar \lambda  + m \I \psi^2_m \sigma^m
   \bar \chi 
 \nonumber \\
& &+\ \sqrt{2} m \I \lambda \chi 
 + {1 \over 2} m \chi \chi 
 + \Red m \Black\, \psi^2_m \sigma^{m n} \psi^2_n  
+ {\kappa \over 2\sqrt{2}}  \epsilon_{i j} \psi^i_m \psi_{n}^{j}
    \cF^{Amn}_{-}   \nn\\
    &&+\ {\kappa \over {2}}  \chi \sigma^m
    \bar \sigma^n \psi^1_m \cD_n \phi
+  {\kappa \over 2 }  \bar \lambda
    \bar \sigma_m \psi^1_n \cF^{Bmn}_{+}
    +
  {\kappa \over {2}} 
    \epsilon^{pqrs}
    \bar \psi^2_{p } \bar \sigma_q \psi^1_r
    \cD_s \phi\nn\\
&&- \I\, {\kappa \over {2}}  \chi \sigma^m
    \bar \sigma^n \psi^1_m v_n 
-  \I \, {\kappa \over {2}} 
    \epsilon^{pqrs}
    \bar \psi^2_{p } \bar \sigma_q \psi^1_r
    v_ss
    \plushc \Bigr)  
\eea
where, as before, $m = \kappa v^2$, and
\bea
\cD_m \phi &\ =\ & \partial_m \phi - {m\over\sqrt{2}} (A_m + B_m) \nonumber \\
\cF^A_{mn} &=& \partial_{[m }A_{n]} + {m\over\sqrt{2}} B_{mn} \nonumber \\
\cF^B_{mn} &=& \partial_{[m }B_{n]} - {m\over\sqrt{2}} B_{mn}\ . 
\eea
This Lagrangian is invariant under an ordinary abelian gauge symmetry,
an antisymmetric tensor gauge symmetry, as well as the following two supersymmetries,
\bea
\delta_{\eta} e^a_m &\ =\ &\I\, \kappa (\eta^i \sigma^a \overline \psi_{m i} + 
    \bar\eta_i \bar\sigma^a \psi_{m}^i) \nonumber \\
\delta_{\eta} \psi^1_m &=& {2\over\kappa}D_m\eta^1 \nonumber \\
\delta_\eta A_m &=& \sqrt{2}\epsilon_{ij}(\psi_m^i\eta^j + \bar\psi_m^i\bar\eta^j) \nonumber \\
\delta_\eta B_m &=& \bar\eta^1\bar\sigma_m\lambda + \bar\lambda\bar\sigma_m\eta^1\nonumber \\
\delta_\eta B_{mn} &=& 2\eta^1\sigma_{mn}\chi 
+ \I\,\eta^1\sigma_{[ m}\bar\psi^2_{n ]}
+ \I\,\eta^2\sigma_{[ m}\bar\psi^1_{n ]}
\plushc \nonumber\\
\delta_\eta \lambda &=& \I\, \hat\cF^B_{mn}\sigma^{mn}\eta^1 - \I \sqrt{2}
v^2 \eta^2\nonumber\\
\delta_\eta \chi &=& \I\, \sigma^m\bar\eta^1 \hat\cD_m\phi - \hat
v_m\sigma^m\bar\eta^1 + 2 v^2 \eta^2\nonumber \\
\delta_{\eta} \psi^2_m &=& {2\over\kappa}
D_m\eta^2 + \I v^2 \sigma_m \bar\eta^2 
- {\I\over\sqrt{2}}\hat\cF^A_{+mn}\sigma^n\bar\eta^1\nonumber \\
   & & +\ \hat\cD_m \phi \eta^1 +\kappa\left( (\bar\psi^1_m\bar\chi)\eta 
- (\bar\chi\bar\eta)\psi^1_m \right) - \I\, \hat v_m\eta^1\nonumber \\
\delta_\eta \phi &=& \chi\eta^1 + \bar\chi\bar\eta^1 
\eea
up to linear order in the fields.
The supercovariant derivatives are given by
\bea
\hat \cD_m\phi&\ =\ & \cD_m\phi - {\kappa\over{2}}(\psi^1_m\chi +
\bar\psi^1_m\bar\chi)  \nonumber\\
\hat \cF^A_{mn}&=& \cF^A_{mn} + {\kappa\over \sqrt{2}}(\psi^2_{
[m}\psi^1_{n] } + \bar\psi^2_{ [m}\bar\psi^1_{n] }) \nonumber \\
\hat \cF^B_{mn}&=& \cF^B_{mn} - {\kappa\over 2}(\bar\lambda\bar\sigma_{
[n}\psi^1_{m ]} + \bar\psi^1_{ [m}\bar\sigma_{n] }\lambda)  \nonumber\\
\hat v_m &=& v_m + \Bigl(\, \I\kappa \psi^{1}_n\sigma_{m}{}^n\chi  
                 -{\I\kappa\over
2}\epsilon_{m}{}^{nrs}\psi_n^{1}\sigma_r\bar\psi_s^{2} 
\plushc \Bigr)  \ .
\eea
These fields form an irreducible representation of the $N=2$
algebra.

Each of the two Lagrangians has a full $N=2$ supersymmetry (up to
the appropriate order).  The first
supersymmetry is realized linearly, so it is not broken.  The second
is realized nonlinearly, so it is spontaneously broken.  In each case,
the transformations imply that
\be
\zeta\ =\ {1\over \sqrt3}\,
(\chi - \I \sqrt 2 \lambda)
\ee
does not shift, while
\be
\nu\ =\ {1\over \sqrt3}\,
(\sqrt 2 \chi + \I \lambda )
\ee
does.  Therefore $\nu$ is the Goldstone fermion for $N=2$ supersymmetry,
spontaneously broken to $N=1$.

In the chiral case, we find
\bea
\left[ \,\delta_{\eta_1}, \,\delta_{\eta_2} \right] \, \phi 
&\ =\ &  2\sqrt{2}\,  v^2\,\eta_1\eta_2  \nonumber \\
\left[ \,\delta_{\eta_1}, \,\delta_{\eta_2} \right] \, {\cal A}_m
&=& {4\over\kappa} \, \partial_m\, \eta_1\eta_2\ .
\eea
The complex scalar $\phi$ is indeed the Goldstone boson
for a gauged central charge.  Moreover, in unitary gauge, where
\be
\phi\ =\ \nu\ =\ 0\ ,
\ee
this Lagrangian reduces to the usual representation for a massive
$N=1$ spin-3/2 multiplet.\cite{fvn}

In the vector case, we have
\bea
\left[ \, \delta_{\eta^2}, \,  \delta_{\eta^1} \right] \,  A_m &\ =\ &
{2\sqrt{2} \over \kappa}
\partial_m(\eta^1\eta^2 + \bar\eta^1\bar\eta^2) 
  - \sqrt{2} \, \I \,  v^2 \, (\eta^2\sigma_m\bar\eta^1 -
\eta^1\sigma_m\bar\eta^2) \nonumber \\
\left[ \,  \delta_{\eta^2}, \,  \delta_{\eta^1} \right] \,  B_m &=& 
\sqrt{2} \, \I \,  v^2 \, (\eta^2\sigma_m\bar\eta^1 -
\eta^1\sigma_m\bar\eta^2) \nonumber \\
\left[ \,  \delta_{\eta^2}, \,  \delta_{\eta^1} \right] \,  B_{mn} &=&
{2 \, \I\over \kappa}D_{[m }
(\eta^2\sigma_{ n]}\bar\eta^1 - \eta^1\sigma_{n] }\bar\eta^2)\ .
\eea
We see that the real vector $-(A_m - B_m)/\sqrt{2}$ is the 
Goldstone boson for a gauged
{\it vectorial} central extension of the $N=2$ algebra.  In addition,
the real scalar $\phi$ is the Goldstone boson associated with a
single real gauged central charge.  In the unitary gauge, with
\be
-{1\over\sqrt{2}}(A_m - B_m)\ =\ \phi\ =\ \nu\ =\ 0\ ,
\ee
this Lagrangian reduces to the alternative representation for
the massive $N=1$ spin-3/2 multiplet.\cite{ogsok}

Now that we have explicit realizations of partial supersymmetry
breaking, we can go
back and see how they avoid the no-go argument presented in the
introduction.  We first compute the second supercurrent.  In each case
it turns out to be
\be
J^2_{m\alpha}\ =\ 
v^2\,(\sqrt6\, \I \,\sigma_{\alpha\dot\alpha m}
\bar\nu^{\dot\alpha} +
4\,\sigma_{\alpha\beta m n} \psi^{2n\beta})
\ee
plus higher-order terms.
Computing, we find
\bea
\{\,\bar Q_{\dot\alpha},\,J^1_{m\alpha}
\,\} &\ =\ & 2\,\sigma^n_{\alpha\dot\alpha}\,
T_{mn}\nonumber \\
\{\,\bar S_{\dot\alpha},\,J^2_{m\alpha}
\,\} &=& 2\,\sigma^n_{\alpha\dot\alpha}\,
T_{mn}\ .
\eea
Now, however, $J^i_{\alpha m}$ and $T_{mn}$ contain contributions
from {\it all} of the fields, including the second gravitino.  When
covariantly-quantized, the second gravitino gives rise to states of
negative norm.  Indeed, it is not hard to check that
\be
(\bar S S + S \bar S)\,
|0\rangle\ =\ 0\ ,
\ee
even though
\be
S\,|0\rangle\ \ne\ 0 \quad\qquad
\bar S\,|0\rangle\ \ne\ 0\ .
\ee
The supergravity couplings exploit the second loophole to the
no-go theorem!

The Lagrangian in the chiral case is a truncation of the supergravity
coupling found by Cecotti, Girardello and Porrati\cite{cgp} and by
Zinov'ev.\cite{zin}
Their results were based on {\it linear} $N=2$ supersymmetry; they
involved $N=2$ vector- and hyper-multiplets.  The Lagrangian for the
vector case is new.  It contains a new realization of $N=2$ supergravity.
In each case, the couplings presented here are minimal and model-independent.
They describe the superHiggs effect in the low-energy effective theories
that arise from partial supersymmetry breaking.

\section{The Massive $N=1$ Spin-3/2 Multiplet in Anti de Sitter Space}

In the last part of this talk, we will examine the question
of partial supersymmetry breaking in anti de Sitter space.  Before
we do this, let us first recall the AdS $N=2$ supersymmetry algebra,
$OSP(2,4)$.  The relevant parts of the algebra are
specified by the following commutators:
\bea
\{ Q^i_\alpha, \bar Q_{j\dot\beta}\}&\ =\ &2\sigma^a_{\alpha\dot\beta}R_a\delta^i_j \nonumber \\
\{ Q^i_\alpha, Q^{\beta j}\}&=&2\I \Lambda {{\sigma^{ab}}_\alpha}^\beta M_{ab}\delta^{ij} +
 2\I {\delta_\alpha}^\beta \epsilon^{ij} T\nonumber \\[1mm]
\left[ T^{ij}, Q^k \right]&=&\I \Lambda (\delta^{jk}Q^i - \delta^{ik}Q^j)\ .
\eea
When the cosmological constant $\Lambda \rightarrow 0$, this contracts the
the usual $N=2$ supersymmetry algebra.  The generator $R_a$ contracts to
the momentum generator $P_a$, while $T$ contracts to a single {\it real} central
charge.  Since the flat-space constructions relied on either a {\it complex}
central charge or a {\it vector} central charge, one is led to wonder
how partial breaking works in AdS space.

In what follows we consider the analog of the chiral case discussed above.
(The vector case is presently under investigation.)  We find it useful
to follow the same procedure as before.
We start with the massive $N=1$ spin-3/2 multiplet in AdS space.  This
multiplet contains the following AdS representations: \cite{nicolai}
\be
D(E+{\textstyle{1\over2}},
{\textstyle{3\over2}})\ \oplus\ D(E,1)
\ \oplus \ 
D(E+1,1)\ \oplus\ 
D(E+{\textstyle{1\over2}},
{\textstyle{1\over2}})
\label{reps}
\ee
where $D(E,s)$ denotes the eigenvalues under $U(1) \times SU(2) \subset
SO(3,2)$ and unitarity requires $E>2$.  ($E$ is the AdS generalization of
the mass.  For this representation, $E \rightarrow 2$ corresponds to the
massless limit.)

The Lagrangian for this multiplet is given by
\bea
e^{-1}\cL &\ =\ & e^{-1}\epsilon^{m n r s} \overline \psi_{m}
  \overline \sigma_n \nabla_r \psi_s
 - \I \overline \zeta \overline \sigma^m \nabla_m \zeta
 - {1 \over 4} A_{m n} A^{m n} - {1 \over 4} B_{m n} B^{m n}
\nonumber \\ & & -\ {1\over 2}(m^2 - m\Lambda) A_m A^m
 - {1\over 2}(m^2 + m\Lambda) B_m B^m  \nonumber \\
& & +\ {1\over 2}m\,\zeta\zeta \ +\ {1\over 2}m\,\bar\zeta\bar\zeta
 -\ m\,\psi_m \sigma^{m n} \psi_n \ -\ m\,\bar\psi_m \bar\sigma^{m n}
\bar\psi_n 
\label{AdSL}
\eea
where $\Lambda \ne 0$ and $\nabla$ denotes the AdS covariant
derivative.  This Lagrangian is invariant under the following
supersymmetry transformations,
\bea
\delta_\eta A_m &=& \sqrt{1 + \epsilon}(\psi_m\eta + \bar\psi_m\bar\eta) \nonumber \\
& & +\ {1\over\sqrt{1 - \epsilon}}
\left( \I{1\over\sqrt{3}}(1 - \epsilon)
(\bar\eta\bar\sigma_m\zeta - \bar\zeta\bar\sigma_m\eta)
-{1\over \sqrt{3}m}\partial_m(\zeta\eta + \bar\zeta\bar\eta)\right) \nonumber \\
\delta_\eta B_m &=& \sqrt{1 - \epsilon}(-\I\psi_m\eta + \I\bar\psi_m\bar\eta) \nonumber \\
& & +\ {1\over\sqrt{1 + \epsilon}}\left( - {1\over\sqrt{3}}(1 +
\epsilon) (\bar\eta\bar\sigma_m\zeta + \bar\zeta\bar\sigma_m\eta)
+{\I\over \sqrt{3}m}\partial_m(\zeta\eta - \bar\zeta\bar\eta) \right) \nonumber\\
\delta_\eta \zeta &=& \sqrt{1 -
\epsilon}({1\over\sqrt{3}}A_{mn}\sigma^{mn}\eta -\I{m\over\sqrt{3}}
\sigma^m\bar\eta A_m) \nonumber \\
& & +\ \sqrt{1 + \epsilon}( - {\I\over\sqrt{3}}B_{mn}\sigma^{mn}\eta
+ {m\over\sqrt{3}}
\sigma^m\bar\eta B_m ) \nonumber \\
\delta_\eta \psi_m &=& {1\over\sqrt{1 + \epsilon}} \left( {1\over
3m}\nabla_m(A_{rs}\sigma^{rs}\eta + 2\I m
\sigma^n\bar\eta A_n) 
 -{\I\over 2}(H^A_{+mn}\sigma^n + {1\over
3}H^A_{-mn}\sigma^n)\bar\eta \right. \nonumber \\ & & \left.
-\ {2\over 3}m({\sigma_m}^n A_n \eta + A_m\eta)
 -{\I\over 2}\epsilon H^A_{+mn}\sigma^n\bar\eta - \epsilon m
A_m\eta \right) \nonumber \\ & & +\ {1\over\sqrt{1 - \epsilon}}\left(
{-\I\over 3m}\nabla_m(B_{rs}\sigma^{rs}\eta - 2\I m 
\sigma^n\bar\eta B_n)
    +{1\over 2}(H^B_{+mn}\sigma^n \right. \nn\\
    &&  +\ {1\over
3}H^B_{-mn}\sigma^n)\bar\eta  \left.
+ {2\over 3}\I m({\sigma_m}^n B_n \eta + B_m\eta)
    -{1\over 2}\epsilon H^B_{+mn}\sigma^n\bar\eta - \I\epsilon m
B_m\eta \right) \ ,\nn\\
\eea
where $\epsilon = \Lambda/m$.  Note that the ``mass" $m$ is defined
to be $m = (E-1)\Lambda$, with $E>2$.  This definition gives masses
consistent with the AdS representations in eq.~(\ref{reps}). The fact
that $E>2$ implies that $0 \le \epsilon \le 1$.

To unHiggs the representation, note that as $E \rightarrow 2$, the
representation (\ref{reps}) splits into
\bea
D({\textstyle{5\over2}},
{\textstyle{3\over2}})\ \oplus\ D(2,1)
\ &&\oplus \ D(3,1)\ \oplus\ 
D({\textstyle{5\over2}},
{\textstyle{1\over2}})\nonumber \\
&&\oplus\ 
D({\textstyle{3\over2}},
{\textstyle{1\over2}})
\ \oplus\ D(3,0)\ .
\eea
These are precisely the degrees of freedom of a massless spin-3/2
multiplet plus a {\it massive} spin-one vector multiplet.  We see
that the resulting symmetry group is actually $OSP(1,4) \times U(1)$,
where the $U(1)$ must be spontaneously broken because $E>2$.  (Usually,
$E \rightarrow 3/2$ is necessary to unHiggs a vector multiplet in AdS
space.  For the case at hand, this would spoil the unitarity of the
spin-3/2 field.)

The steps to derive the unHiggsed Lagrangian are very similar to
those for the chiral multiplet in flat space.   We find
\bea
& &e^{-1}\cL \ =\ \nonumber\\
&  & -\ {1 \over 2
\kappa^2} {\cal R} + \epsilon^{m n r
s}
\overline
\psi_{i m}
  \overline \sigma_n D_r \psi^i_s 
 - \I \overline \lambda \overline \sigma^m D_m \lambda  
 - \I \overline \chi \overline \sigma^m D_m \chi  \nonumber \\
& & -\ {1 \over 4} A_{m n} A^{m n} - {1 \over 4}
B_{m n} B^{m n} - {1\over 2}\cD_m \phi_A \cD^m
\phi_A 
 - {1\over 2}\cD_m \phi_B \cD^m \phi_B   \nonumber \\
& & - \ \Bigl( {1\over \sqrt{2}} m \sqrt{1 - \epsilon^2}
\psi^2_m \sigma^m
   \overline \lambda 
+ m \sqrt{1 - \epsilon^2}  \I\psi^2_m \sigma^m
   \overline \chi  \nonumber \\
& &+\ \sqrt{2} m \I\lambda\chi
   + {1\over 2}m\chi\chi 
\Red +\ m\psi^2_m \sigma^{m n} \psi^2_n 
-
     \epsilon m\psi^1_m \sigma^{m n} \psi^1_n \Black \nonumber \\
& &+\ {\kappa \over 4}  \epsilon_{i j} \psi^i_m
\psi_{n}^{j}
    (\sqrt{1 + \epsilon}H_{A-}^{m n} - \I \sqrt{1 - \epsilon}H_{B-}^{m
n})  \nonumber \\
& & +\ {\kappa \over {2}}  \chi \sigma^m
    \overline \sigma^n \psi^1_m (\cD_n \phi_A - \I \cD_n \phi_B)
 \nonumber \\
& &+\ {\kappa \over 2 \sqrt{2}}  \overline \lambda
    \overline \sigma_m \psi^1_n (\sqrt{1 - \epsilon}H_{A+}^{m n} - \I
\sqrt{1 + \epsilon}H_{B+}^{m n})
    \nonumber \\
& & +\  {\kappa \over {2}} 
    \epsilon^{m n r s}\sqrt{1 - \epsilon\over 1 +
\epsilon} 
    \overline \psi_{m 2} \overline \sigma_n \psi^1_r
    (\partial_s \phi_A - \I \partial_s
\phi_B)
    \nonumber \\
& & -\ {\kappa \over {2}}m 
    \epsilon^{m n r s} 
    \overline \psi_{m 2} \overline \sigma_n \psi^1_r
    (\sqrt{1 + \epsilon}A_s - \I\sqrt{1 -
\epsilon}B_s)
     \nonumber \\
& & -\ 2\kappa\epsilon m \sqrt{1 - \epsilon\over
1 + \epsilon} \bar\psi_{m 2}
\bar\sigma^{mn}\bar\psi_{n 1}\phi_A
+ {\kappa\epsilon m \over
\sqrt{2}}\bar\lambda\bar\sigma^m\psi^{1}_m \phi_A
  \nonumber \\
& & +\   \I {\kappa\epsilon
m}\bar\chi\bar\sigma^m\psi^{1}_m \phi_A \plushc \Bigr)
\label{lagr} + 3{\epsilon^2
m^2\over \kappa^2} \ .
\label{AdSLG}
\eea
This Lagrangian is invariant (to lowest order in the fields)
under the following supersymmetry transformations,
\bea
\delta e^a_m &\ =\ &\I \,  \kappa \eta^i \sigma^a \overline \psi_{m i} + 
   \I \,  \kappa \bar\eta_i \bar\sigma^a \psi_{m}^i \nonumber \\
\delta_{\eta} \psi^1_m &=& {2\over\kappa}D_m\eta^1 + \I \,  {\epsilon
m\over \kappa} \sigma_m \bar\eta^1 \nonumber \\
\delta_\eta A_m &=& \sqrt{1 + \epsilon}\epsilon_{ij} (\psi_m^i\eta^j +
\bar\psi_m^i\bar\eta^j) + \sqrt{1 - \epsilon}
{1\over\sqrt{2}}
(\bar\eta^1\bar\sigma_m\lambda + \bar\lambda\bar\sigma_m\eta^1)
\nonumber \\
\delta_\eta B_m &=& \sqrt{1 - \epsilon}\epsilon_{ij} (-\I \,
\psi_m^i\eta^j + \I \, \bar\psi_m^i\bar\eta^j) + 
\sqrt{1 + \epsilon}{\I\over\sqrt{2}}
(\bar\eta^1\bar\sigma_m\lambda - \bar\lambda\bar\sigma_m\eta^1) 
\nonumber\\
\delta_\eta \lambda &=& \I \, \sqrt{1 - \epsilon}{1\over\sqrt{2}}\hat
A_{mn}\sigma^{mn}\eta^1 + \sqrt{1 + \epsilon}
{1\over\sqrt{2}}\hat B_{mn}\sigma^{mn}\eta^1 \nonumber \\
& & + \sqrt{2} \, \I \, \epsilon m  \, \phi_A\eta^1 
 -\I \, \sqrt{2} \,  {m\over \kappa} \sqrt{1 - \epsilon^2} \eta^2
\nonumber \\
\delta_{\eta} \chi &=& \I \, \sigma^m\bar\eta^1 \hat\cD_m\phi_A -
\sigma^m\bar\eta^1 \hat\cD_m\phi_B 
 - 2 \, \epsilon m \,  \phi_A\eta^1  
+ 2 {m\over \kappa} \sqrt{1 - \epsilon^2} \eta^2 \nonumber \\
\delta_\eta \psi^2_m &=& {2\over\kappa}D_m\eta^2 + \I \,  {m\over
\kappa} \sigma_m \bar\eta^2 
-{\I\over 2}\sqrt{1 + \epsilon} \hat H^A_{+mn}\sigma^n\bar\eta^1 -
m\sqrt{1 + \epsilon}A_m\eta^1 \nonumber \\
& & + {1\over 2}\sqrt{1 - \epsilon}\hat H^B_{+mn}\sigma^n\bar\eta^1 +
\sqrt{1 - \epsilon\over 1 + \epsilon}( \partial_m \phi_A - 
\I \, \cD_m\phi_B)\eta^1   \nonumber \\
& & -{\kappa\over 2}\sqrt{1 - \epsilon\over 1 + \epsilon}\psi^1_m
(\delta_{\eta^1}\phi_A 
 -\I \, \delta_{\eta^1}\phi_B) 
-\I \, \epsilon m\sqrt{1 - \epsilon\over 1 + \epsilon}  \phi_A
\sigma_m \bar\eta^1 \nonumber \\
\delta_\eta \phi_A &=& \chi\eta^1 + \bar\chi\bar\eta^1 \nonumber \\ 
\delta_\eta \phi_B &=& -\I \, \chi\eta^1+ \I \, \bar\chi\bar\eta^1
\label{AdST}
\eea
where
\bea
\cD_m \phi_A &=& \partial_m \phi_A - m\sqrt{1 - \epsilon}A_m \nonumber \\
\cD_m \phi_B &=& \partial_m \phi_B - m\sqrt{1 + \epsilon}B_m 
\eea
and
\bea
\hat \cD_m\phi_A&=& \partial_m\phi_A - m\sqrt{1 - \epsilon}  A_m  - {\kappa\over{2}}(\psi^1_m\chi + \bar\psi^1_m\bar\chi)  \nonumber\\
\hat \cD_m\phi_B&=& \partial_m\phi_B - m\sqrt{1 + \epsilon}  B_m + \I {\kappa\over{2}}(\psi^1_m\chi - \bar\psi^1_m\bar\chi) \nonumber\\
\hat A_{mn}&=& A_{mn} + {\kappa\over 2}\sqrt{1 + \epsilon}(\psi^2_{ [m}\psi^1_{n] } + \bar\psi^2_{ [m}\bar\psi^1_{n] }) \nonumber \\
& &- \sqrt{1 - \epsilon}{\kappa\over2\sqrt{2}}(\bar\lambda\bar\sigma_{
[n}\psi^1_{m ]} + \bar\psi^1_{ [m}\bar\sigma_{n] }\lambda)  \nonumber
\\
\hat B_{mn}&=& B_{mn} - \I {\kappa\over 2}\sqrt{1 - \epsilon}(\psi^2_{ [m}\psi^1_{n] } - \bar\psi^2_{ [m}\bar\psi^1_{n] }) \nonumber \\
& &+ \I\sqrt{1 + \epsilon}{\kappa\over2\sqrt{2}}(\bar\lambda\bar\sigma_{
[n}\psi^1_{m ]} - \bar\psi^1_{ [m}\bar\sigma_{n] }\lambda)  \ .
\eea
Note that this Lagrangian depends on $\kappa,$ $v$ and $\Lambda$, with
$m = \sqrt{\Lambda^2 + \kappa^2 v^4}$ and $\epsilon = \Lambda/m$.  As
before, $0 \le \epsilon \le 1$.

The Lagrangian (\ref{AdSLG}) describes the spontaneous breaking of $N=2$
supersymmetry in AdS space.  It has $N=2$ supersymmetry and a local $U(1)$
gauge symmetry.  In unitary gauge, it reduces to the massive $N=1$
Lagrangian of eq.~(\ref{AdSL}).

\begin{figure}[t]
\hspace*{0.6truein}
\psfig{figure=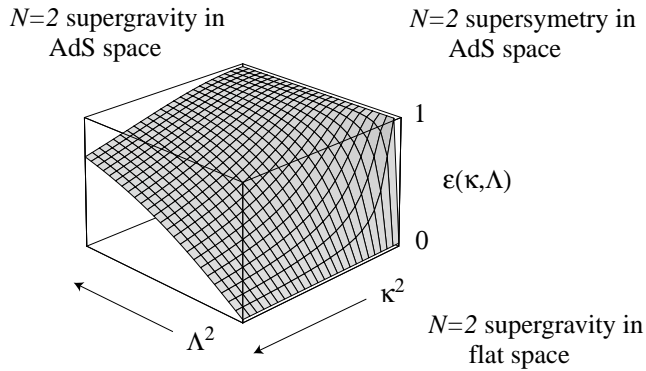,height=2.0in}
\caption{The manifold of partially-broken $N=2$ supergravity theories as a function
of Newton's constant $\kappa$ and the cosmological constant $\Lambda$.}
\label{chiral}
\end{figure}

In it instructive to consider the Lagrangian (\ref{AdSLG}) in two limits.
The first, with $\Lambda \rightarrow 0$, but fixed $\kappa$ and
$v$, corresponds to the case $\epsilon \rightarrow 0$.  The Lagrangian
(\ref{AdSLG}) reduces to the previous case, of partially broken
supersymmetry, coupled to supergravity, in a flat Minkowski background.
The second limit, with $\kappa \rightarrow 0$, but fixed $v$ and
$\Lambda$, corresponds to $\epsilon \rightarrow 1$.  In this limit,
the Lagrangian (\ref{AdSLG}) describes partially broken $N=2$ supersymmetry
in a fixed AdS background.
The full manifold of the $N=2$ supergravities is presented in Figure 2,
where $\epsilon$ is plotted as a function of $\kappa$ and
$\Lambda$.  The two limits correspond to edges of the plot.  The
center region contains the new AdS supergravity presented here.

To find the algebra associated with the partial supersymmetry
breaking, let us consider the second limit, with $\kappa \rightarrow
0$ and fixed $v$, $\Lambda$.  A simple computation shows
that
\bea
\left[\, \delta_{\eta^2},\, \delta_{\eta^1} \,\right] \phi_A &\ =\ & 
2v^2 (\eta^1\eta^2 + 
\bar\eta_1\bar\eta_2) \nonumber \\
\left[\, \delta_{\eta^2},\, \delta_{\eta^1}\, \right] A_m &=& 0 
\label{A}
\eea
while
\bea
\left[\, \delta_{\eta^2},\, \delta_{\eta^1} \,\right] \phi_B &\ =\ & -2 \I v^2 (\eta^1\eta^2 - 
\bar\eta_1\bar\eta_2) \nonumber \\
\left[\, \delta_{\eta^2},\, \delta_{\eta^1} \,\right] B_m &=&
-\sqrt{2} \I v^2 \, {\partial_m\over m}\,(\eta^1\eta^2 - 
\bar\eta_1\bar\eta_2) \ .
\label{B}
\eea
Equation (\ref{A}) tells us that the real scalar $\phi_A$ is the
Goldstone boson associated with the $U(1)$ generator of the
AdS algebra.  (It is this generator which contracts to a real
central charge in flat space.)  Equation (\ref{B}) indicates that
the scalar $\phi_B$ is also a Goldstone boson associated with a
spontaneously-broken $U(1)$ symmetry, but one which is gauged by
the vector field $B_m$.

These results imply that when $v$, $\Lambda \ne 0$, the full
current algebra is actually $OSP(2,4) \times_s U(1)$, nonlinearly
realized.  The $\times_s$ denotes a semi-direct product, because
\Ref{B} shows that the supersymmetry generators close into the
local $U(1)$ symmetry.  Note that this construction evades the
constraints of the Coleman-Mandula/Haag-{\L}opusza\'nski-Sohnius theorem
because the broken supercharges do not exist.  The $OSP(2,4) \times_s
U(1)$ symmetry only exists at the level of the current algebra.
The $U(1)$ symmetry is always spontaneously broken.

\section{Summary}

In this talk we have examined the partial breaking of supersymmetry
flat and anti de Sitter space.  We have seen that partial breaking
in flat space can be accomplished using either of two representations
of the massive $N=1$ spin-3/2 multiplet.  We unHiggsed each
representation, and found a new $N=2$ supergravity and a new $N=2$
supersymmetry algebra.  We also saw that the partial supersymmetry
breaking in AdS space can give rise to a new $N=2$ supersymmetry
algebra, albeit one that is necessarily nonlinearly realized.

\section*{Acknowledgements}
We would like to thank A.~Galperin and S.~Osofsky for enjoyable
collaboration.  This work was supported by the National Science
Foundation, grant NSF-PHY-9404057.

\section*{References}


\bibliography{suhi}
\bibliographystyle{unsrt}


\end{document}